\documentclass[conference]{IEEEtran}
\IEEEoverridecommandlockouts
\usepackage{cite}
\usepackage{amsmath,amssymb,amsfonts}
\usepackage{algorithmic}
\usepackage{graphicx}
\usepackage{textcomp}
\usepackage{xcolor}
\usepackage{tabularx}
\usepackage{booktabs}
\usepackage{multicol}   
\def\BibTeX{{\rm B\kern-.05em{\sc i\kern-.025em b}\kern-.08em
    T\kern-.1667em\lower.7ex\hbox{E}\kern-.125emX}}
\usepackage[english]{babel}
\newcommand*{\Dr}{%
    \iflanguage{english}{Dr.~}{%
    \iflanguage{english}{D\textsuperscript{r}~}{}%
        }}
\usepackage{graphicx}
\usepackage{mathtools}
\usepackage{lipsum}

\usepackage{subcaption}
\usepackage[para,online,flushleft]{threeparttable}
\usepackage{booktabs}
\usepackage{graphicx}
\usepackage{paralist}
\usepackage{hyperref}
\usepackage{comment} 
\usepackage{titlesec} 
\usepackage{float}
\usepackage{dblfloatfix}

\hypersetup{final}
\titlespacing{\subsection}{0pt}{*1}{*1}
\makeatletter
\newcommand{\linebreakand}{%
  \end{@IEEEauthorhalign}
  \hfill\mbox{}\par
  \mbox{}\hfill\begin{@IEEEauthorhalign}
}
\makeatother
\begin{document}

\title{Network Calculus Boosts Open World Learning Capability for Graph Convolution in Routing Networks\\}
\author{
    \IEEEauthorblockN{Yifei Jin}
    \IEEEauthorblockA{
        \textit{KTH Royal Institute of Technology \& Ericsson AB} \\
        \textit{Division of Theoretical and Computer Science \& Ericsson Research}\\
        Stockholm, Sweden \\
        \url{yifeij@kth.se}
    }
\and
    \IEEEauthorblockN{Marios Daoutis}
    \IEEEauthorblockA{
    \textit{Ericsson AB} \\
    \textit{Ericsson Research}\\
        Stockholm, Sweden \\
        marios.daoutis@ericsson.com
    }
\linebreakand 
    \IEEEauthorblockN{Sarunas Girdzijauskas}
    \IEEEauthorblockA{
        \textit{KTH Royal Institute of Technology} \\
        \textit{Division of Software and Computer Systems}\\
        Stockholm, Sweden \\
        \url{sarunasg@kth.se}
        }
\and
    \IEEEauthorblockN{Aristides Gionis}
    \IEEEauthorblockA{
        \textit{KTH Royal Institute of Technology} \\
        \textit{Division of Theoretical and Computer Science}\\
        Stockholm, Sweden \\
        \url{argioni@kth.se}
        }
}
\maketitle
\begin{abstract}
Accurate routing network status estimation is a key component in Software Defined Networking. However, existing deep-learning-based methods for modeling network routing are not able to extrapolate towards unseen feature distributions. Nor are they able to handle scaled and drifted network attributes in test sets that include open-world inputs. To deal with these challenges, we propose a novel approach for modeling network routing, using Graph Neural Networks. Our method can also be used for network-latency estimation. Supported by a domain-knowledge-assisted graph formulation, our model shares a stable performance across different network sizes and configurations of routing networks, while at the same time being able to extrapolate towards unseen sizes, configurations, and user behavior. We show that our model outperforms most conventional deep-learning-based models, in terms of prediction accuracy, computational resources, inference speed, as well as ability to generalize towards open-world input.
\end{abstract}


\section{Introduction}
\label{sec:intro}
\begin{figure}[t!]
    \centering
    \includegraphics[width=0.5\textwidth]{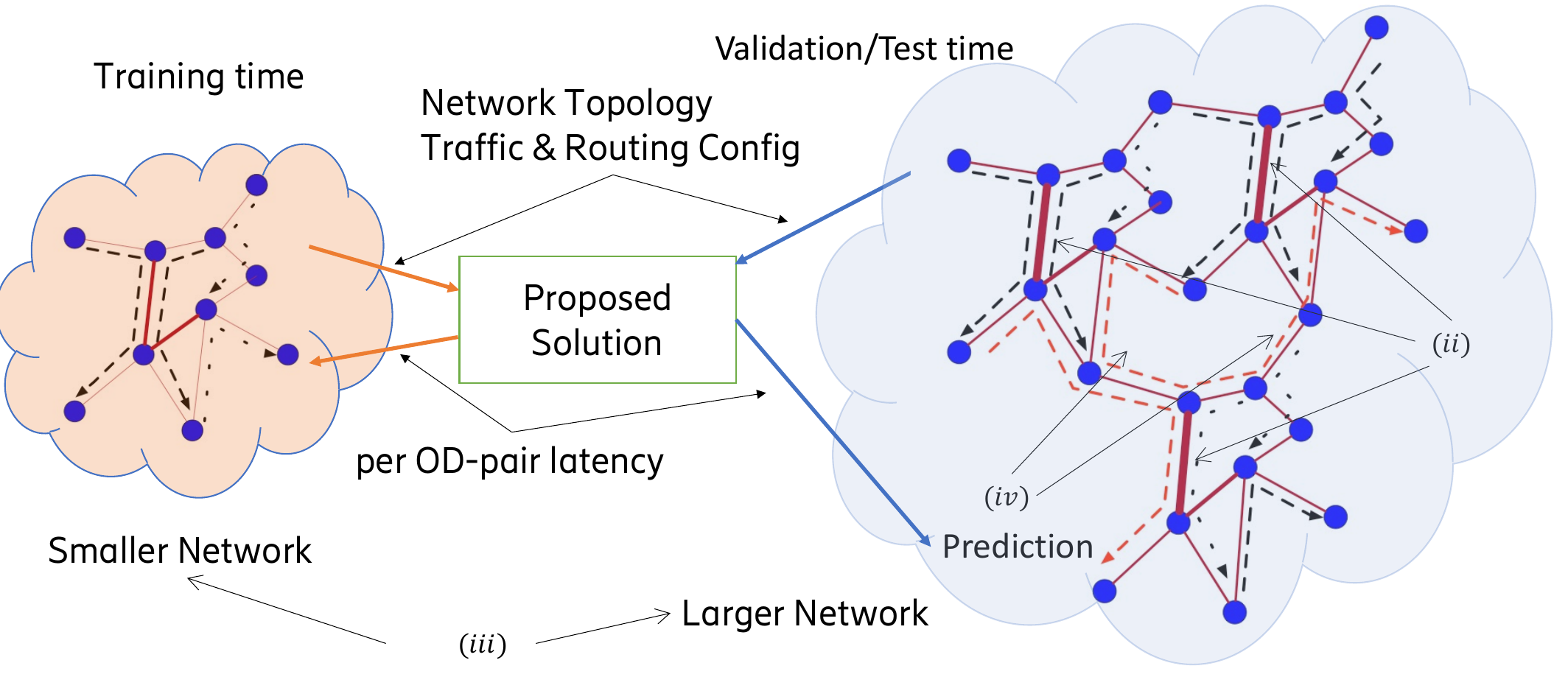}
    \caption{Schematic Representation of SDN Network State Estimation Problem. Directed line denotes network traffic flows with different throughput in the given topology, where one can see that there exist different distributions of line type between training and validation/test time, as mentioned in $(i)$. Undirected line denote the link communicating pair of network elements, with the line's strength denoting the link's capacity level.}
    \label{fig:formulation}
\end{figure}

Software-defined networking (SDN) is a state-of-the-art approach to network management, which permits dynamic and programmatic network configurations (e.g., routing), aimed at improved network performance and monitoring, abstracted away from individual network elements into a centralized network control layer. Consequently, the orchestration and SDN control software is expected to operate on and adapt to unseen network deployments and topologies.
Precise routing-network modeling is a prerequisite for SDN to efficiently estimate and forecast network state.
As a result, in certain scenarios, such as during network expansion, it is expected that: \begin{inparaenum}[\itshape (i)\upshape] 
\item several network features (possibly used in model training) become greatly imbalanced;
\item unseen attribute values are expected to be out-of-distribution;
\item varying, often larger, network sizes are encountered; and 
\item relational dependencies expand among longer node chains.
\end{inparaenum}
An illustrative example of above is shown in Fig.~\ref{fig:formulation}.
A successful approach to deal with the above-mentioned challenges, is to employ ideas from the domain of open-world machine learning (OWL)~\cite{parmar2021open}. Compared with traditional machine learning, OWL is expected to extrapolate towards unseen input distribution, classes, and scaling in the training set.
Note, that the problem of estimating the network state is \hbox{EXPTIME}-complete, as shown by~\cite{Gamarnik2011AlgorithmicCI}, and has been extensively studied in the literature~\cite{amin2021survey}, where most commonly conventional deep-learning methods are employed in the context of reinforcement learning (RL) based network orchestration as well as deep-learning-based service optimization. 

In this paper we consider the problem of estimating the latency between a source and a destination network node (denoted as {\em OD} pair) in a routing network. 
The problem setting is to learn a model on smaller networks and extrapolate the predictions on larger networks assuming open-world input, as illustrated in Fig.~\ref{fig:formulation}.

Conventional deep learning (DL) methods have focused in predictive user behavior modeling on service demand and data usage. In this setting, a line of work focuses on leveraging deep neural networks for adaptive optimisation of routing schemes, as by~\cite{srivastava2021machine}, which typically require similar data distribution in the test environment. Only few papers have considered to extrapolate their proposed model on a larger network~\cite{ferriol2021scaling}, something which, this far, has been considered infeasible in real-world scenarios. 
Moreover, successful extrapolation would require introducing task-specific non-linearity (i.e., queuing theory and network calculus in the routing network~\cite{ciucu2012perspectives}) in the model~\cite{xu2020neural}, something which, this far, has not been considered in state-of-the-art DL solutions.

Graph Neural Networks (GNNs) have been used successfully for incorporating domain knowledge into different problem settings, which in turn, help to devise robust models better suited for OWL in learning complex-system representations. 
Moreover, our work is primarily inspired by~\cite{sun2019sinet} who reveal that by solely observing key nodes in the SDN topology, the deep network can produce effective embeddings capable of predicting network performance.

Our study is among the first approaches that attempt to model the routing network state snap\-shot by learning the given topology structure for the task of estimating the network latency using open-world input.
Our proposed solution transforms the task of estimating the latency between source and destination nodes, to a \emph{link-attribute prediction task}, i.e., predicting each link that can potentially contribute to traffic delays occurring in the routing trajectory. Estimating every link's latency contribution requires three aspects:
($i$) the traffic amount passing through the link;
($ii$) the capacity of the link; and 
($iii$) link's structural features.
The link traffic and capacity jointly denote the data throughput,
and the congestion state of the given link. 
The link's structural features 
can be learned from the network topology and they serve two purposes: 
($i$) to identify key links that can become bottleneck 
of the network performance; and 
($ii$) to compute pair-wise similarities, 
which are used to discover similar links.
We train our model on distributions of a particular network size, yet, the model achieves at least $75\%$ reduction in mean absolute percentage error (MAPE) measured in different sized networks during inference, i.e., without explicit training using the larger (up to 6$\times$) network topologies. Last but not least, our solution requires no GPU resource while still ensures a faster training as well as $3$--$7$ times faster inference speeds, with only 6-25\% embedding size required, when compared, for example, with the best-performing benchmark models.
The remaining sections are organised as follows: In Sec.~\ref{sec:related} we discuss the related approaches in the domain of size generalisation of GNNs and graph-based ML in SDN routing, while in Sec.~\ref{sec:formulation} we present our problem formulation in more detail. In Sec.~\ref{sec:method} we present our methodology followed by Sec.~\ref{sec:exp}, which contains the experimental evaluation of our algorithm. Finally Sec.~\ref{sec:conclusion} concludes with a discussion of our main contributions and considerations for future work.

\section{Related Work}
\label{sec:related}
Graph Convolutional Networks (GCN) is a known feature-extraction technique for non-Euclidean spaces~\cite{kipf2016semi}. GCNs are capable of learning the graph structure by aggregating each node's embedding with its neighbors. In the conventional architecture, GCN is proven to be capable of incorporating node and edge information in a undirected graph scope. We are currently witnessing an increasing number of publications on GCN models, which, on the one hand reveal the success of this model, and on the other hand expose its limitations.

GCN models are constrained due to aggregating only local neighbors. For large enough graphs, GCN may fail to capture distant (long-range) dependencies entirely. Efforts have been made towards both building deeper GNN models~\cite{li2021training} and building expressive GNN layers on long-range dependencies. For example, \cite{chen2019path} proposed the first effective method to encode long-range dependencies, between any two nodes, of the graph in the node embedding. 
Furthermore,~\cite{yang2021spagan} propose to apply graph attention network (GAT), to compute shortest-path-to-node attention, based on a transformer~\cite{velivckovic2017graph} architecture. 
This work is one of the most related approaches to the one presented here. Contrary to the aforementioned contributions, which are based on transductive settings, our work focuses instead on an inductive setting. In our model, we are incorporating path-to-node attention, but passively between the path's structurally similar node and the targeting node, to tackle longer chains of dependencies in the test data. In addition, we do not build very deep GNN solutions to enlarge the receptive field up to the length of dependency, for balancing between model accuracy, resource usage and inference speed trade-off.

In a survey by~\cite{xu2020neural}, the extrapolation ability of GNNs regards mainly two aspects: the ability to extrapolate towards \emph{unseen structural features} as well as to extrapolate towards \emph{out-of-distribution attributes}. \cite{yehudai2021local} conclude that the size generalization can only be applied on graphs with certain structural features. Moreover, they suggest that the expressiveness of the GNN model must be still sufficient for applying it to larger graphs. \cite{balcilar2020analyzing} further elaborate from the perspective frequency response, pointing out that spectral graph convolution is inherently transferable for graphs with similar degree distribution. 

In our model, we leverage the conclusions from the aforementioned works and cast the graph-size-generalization problem to a transferred formulation of graph,
which is considered to be learnable by spectral graph-convolution methods. On the other hand, extrapolation towards out-of-distribution data is a fundamental challenge for most neural networks. For out-of-distribution attributes, the extrapolation ability relies more on the embedding and readout function of the GNN architecture. \cite{xu2020neural} have shown the multi-layered perception models (MLP) can only extrapolate if the test set is expanded from the training set in all geometrical directions in the latent space. Both~\cite{trask2018neural} and~\cite{madsen2020neural} have studied on approximating simple arithmetic computation ($+$, $-$, $\times$, and $/$) when extrapolate towards drifted numerical values. In our proposed model, we utilize building blocks from \cite{trask2018neural} to encourage the model to learn the arithmetic impact, when tackling drifted numerical input features. 


 

More recent related works focus on learning the tuned graph representation of routing networks. \cite{rusek2020routenet} propose RouteNet, which is regarded as a first work that introduces message passing in deep-learning-based network modeling, in a bipartite graph formulation.
\cite{happ2021graph} and \cite{kong2021path} extend RouteNet to be adaptive to heterogeneous scheduling policies and routing scheme, while improving its accuracy and inference speed on similar topology structures.
Finally,~\cite{ferriol2021scaling} present their work on topology size generalization for latency estimation of Origin-Destination (OD) pairs, the same problem we focus here. Through improving RouteNet by including queue occupancy state per link, beside the path and link nodes, they formulate a tri\-partite message passing scheme, which introduces size generalization ability to the model. 
Yet, none of aforementioned works utilize the topology's structural feature in network-state embedding, as we address in this work.
\section{Problem Formulation}
\label{sec:formulation}

The task we consider in this paper is to perform \emph{mean latency estimation} on every OD pair $p$ in the network. 
We summarize the defined notation in Table.~\ref{table:notation}. The input to our problem is the following.

\begin{enumerate}
    \item 
    A sequence of attributed size-varying networks $\mathbf{G} = \{G_i(V_i,E_i;X_i,Y_i)\}$,
    where $V_i$ denotes the $|V_i| = n_i$ network nodes and 
    $E_i$ denotes the $|E_i| = m_i$ links between pairs of nodes in $V_i$. 
    For each node $v \in V_i$, a set of node attributes $x(v)$ are provided to describe the heterogeneous routing configuration on the network node level. Similarly, a set of link attributes $y(e)$ is introduced for each link $e\in E_i$ to denote link configuration attributes with respect to routing.
    The set of all node attributes in $G_i$ is denoted by $X_i$, 
    and the set of all link attributes is denoted by $Y_i$. 
    Each network $G_i$ includes $k_i$ OD pairs $P_i=\{p_{i,j}\}$, 
    with $j=1,\ldots,k_i$.
    \item 
    A sequence of traffic matrices $\mathbf{T} =\{T_i\}$, where each $T_i$ corresponds to a network $G_i$. 
    Each traffic matrix $T_i$ consists by $k_i$ vectors $\{t_{i,j}\}$, 
    with $j=1,\ldots,k_i$, 
    where each $t_{i,j} \in \mathbb{R}^2$ denotes the mean throughput and peak throughput features
    for the OD pair $p_{i,j}$.
    Note that there exist OD pairs with same source and destination nodes but different throughput features. 
    \item 
    A sequence of routing matrices $\mathbf{R} =\{R_i\}$, 
    where each $R_i$ with dimension $\mathbb{R}^{n_i \times n_i}$, with $0$'s on the diagonal elements, corresponds to network $G_i$.
    Each element in $R_i$, 
    denoted by $R_{i(c,d)} = r_i({v_{c}},v_{d}); v_c,v_d \in V_i$ 
    represents the next-hop per OD pair, 
    given the current node $v_{c}$ and the destination node $v_{d}$. 
    Note that $R_i$ contains information for all possible routing pairs,
    regardless of the given OD pair elements in $T_i$. 
    \item 
    \emph{The ground truth}: 
    A sequence of performance matrices $\mathbf{Q} = \{Q_i\}$, where each $Q_i$ corresponds to network $G_i$. Each performance matrix $Q_i$ consists of $k_i$ vectors $\{q_{i,j}\}$, with $j=1,\ldots,k_i$, denoting the performance of each OD pair $p_{i,j}$ in global (mean latency) and local (per-link occupancy) metrics. The task is to estimate the mean latency value of each OD pair, denote as an element in $q_{i,j} = Q_i(p_{i,j})$, with $j=1,\ldots,k_i$.
\end{enumerate}
Previous works~\cite{ferriol2021scaling,rusek2020routenet,kong2021path,happ2021graph} formulate the problem defined above as a multi-step prediction problem. More specifically, given network snapshots $(G_i,T_i,R_i)$, the goal is to predict the future $\Delta$ steps of the network state $[Q_i'^1,\ldots,Q_i'^\Delta]$ and pool the result of each step for a prediction $Q_i'$. In contrast, in our model, we consider this problem as a node attribute prediction problem. 
We introduce a modified problem formulation:
\begin{enumerate}
\item Given the input to our problem specified by a sequence of tuples $(G_i(V_i,E_i;X_i,Y_i),T_i,R_i)$, we first transform $G_i(V_i,E_i;X_i,Y_i)$ to a directed graph, where each undirected edge $(u,v) \in E_i$ is replaced by two directed edges $(u,v)$ and $(v,u)$, and then transform it to a line graph $G^{\mathcal{L}}_i(V^{\mathcal{L}}_i,E^{\mathcal{L}}_i)$. We define a projection $\mathcal{L}^{-1}$, to project the representation (i.e: $v_i \in V^{\mathcal{L}}_i,e^{\mathcal{L}}_i \in E^{\mathcal{L}}_i$) in line graph $G^{\mathcal{L}}_i$, back to its representation in $G_i$,
\item The edge set $E^{\mathcal{L}}_i$ of the line graph $G^{\mathcal{L}}_i$ is directed. The node set $V_i^{\mathcal{L}}$ includes only valid routings in $R_i$, i.e., for all $v \in V_i^{\mathcal{L}}$ its respective edge representation $\mathcal{L}^{-1}(v) = e(v_x,v_y)$ is in $E_i$. Thus, there exists $v_d \in V_i$ such that  $r_i(v_x,v_d) = v_y$ holds.
\item We can compute the summed traffic $T^{s}_i$ per-link over all OD trajectory pairs passing through, for all $n_i$ nodes. Each element is denoted by $t^{s}(i,v) = T^{s}_i(v)$, for all $v \in V^{\mathcal{L}}_{i}$. Given node $v$'s edge representation $\mathcal{L}^{-1}(v) = e_v \in E_i$, We define the node attributes $X^{\mathcal{L}}_i(v)$ in $G^{\mathcal{L}}_i$ as follows:
\begin{align}
\label{equ:nodeattr}
\begin{split}
X^{\mathcal{L}}_i(v) = \frac{t^{s}(i,v)}{c(Y_i(e_v))}
\text{ and } t^{s}(i,v) = \sum_{p_k;v \in p_k}T_i(p_k),&\\
\text{ for } v \in V^{\mathcal{L}}_i \text{ and } e_v \in E_i \text{ and } \mathcal{L}^{-1}(v) = e_v&
\end{split}
\end{align}
where $c(Y_i(e_v)) \in Y_i(E_i)$, denotes the capacity per-link in $G_i$. 
\item For each edge $e \in E^{\mathcal{L}}_i$ connecting node $v_s$ to node $v_d$ in $G^{\mathcal{L}}_i$. We introduce corresponding edge weight $w_i(v_s,v_d)$ as follows:
\begin{align}
\label{equ:edgeweight}
\begin{split}
    w_i(v_s,v_d) = 
    \frac{\sum_{p_k;v_s,v_d \in p_k}T_i(p_k)}{t^{s}(i,v_s)}  
    \cdot\frac{ Y_i(e_{v_s})}{Y_i(e_{v_d})},&\\
    \text{ for } 
    v_s,v_d \in V^{\mathcal{L}}_i
    \text{ and }
    e_{v_s},e_{v_d} \in E_i,&\\
    \text{ and }
    \mathcal{L}^{-1}(v_s) = e_{v_s}, \mathcal{L}^{-1}(v_d) = e_{v_d},&
\end{split}
\end{align}
\end{enumerate}
The motivation behind this re-formulation is multi-faceted: \begin{inparaenum}[\itshape (i)\upshape] \item 
Using directed links in $G^{\mathcal{L}}_i$ introduces OD trajectory pairs information into the graph structure. 
The new formulation separates the link's adjacency into in-degree and out-degree, which will be used to represent the link's in-coming traffic (in-traffic) and out-going traffic (out-traffic). 
\item For the task of estimating path latency $Q_i(p_{i,j})$, the problem can be decomposed into estimating the 
latency contribution of each hop 
$v \in V^{\mathcal{L}}_i$ into $\sum_{v;v \in p_k} Q_i(v)$. 
According to queuing theory, such latency contribution
further reduces to estimating $v$'s queue occupancy state $O(i,v) \in Q_i$. Thus, in the graph $G^{\mathcal{L}}_i$, the problem is reduced to a node attribute prediction task.\footnote{A detailed transform procedure is shown in Appendix.A. Link to the appendix: \url{https://github.com/bluelancer/RoutingGNN_Appendix/blob/main/IEEE_Graph_Covolution_for_Routing__Appendix__.pdf}} 
\item Inspired by previous work, identifying key link that is shared by many OD pairs could have a dominating impact on the whole network performance. In the graph $G^{\mathcal{L}}_i$, such feature is disclosed within the ego-networks of nodes $v \in V^{\mathcal{L}}_i$, 
which can be jointly formed by the trajectories of all paths passing through $v$.
Thus, one can discover the importance of node $v$ in network topology by its {\em role} in $G^{\mathcal{L}}_i$. \item The features computed in Eq.~\ref{equ:nodeattr} and Eq.~\ref{equ:edgeweight} are supported by queuing theory;\footnote{A detailed proof is given Appendix.B-C, Link to the appendix same as above.} 
here we consider domain knowledge to be the key in producing a size-invariant graph signal on $G^{\mathcal{L}}_i$. An illustrative example of re-formulation, comparing with the previous works' is given in Fig.~\ref{fig:comparative}.
\begin{figure}
    \centering
    \includegraphics[width=0.4\textwidth]{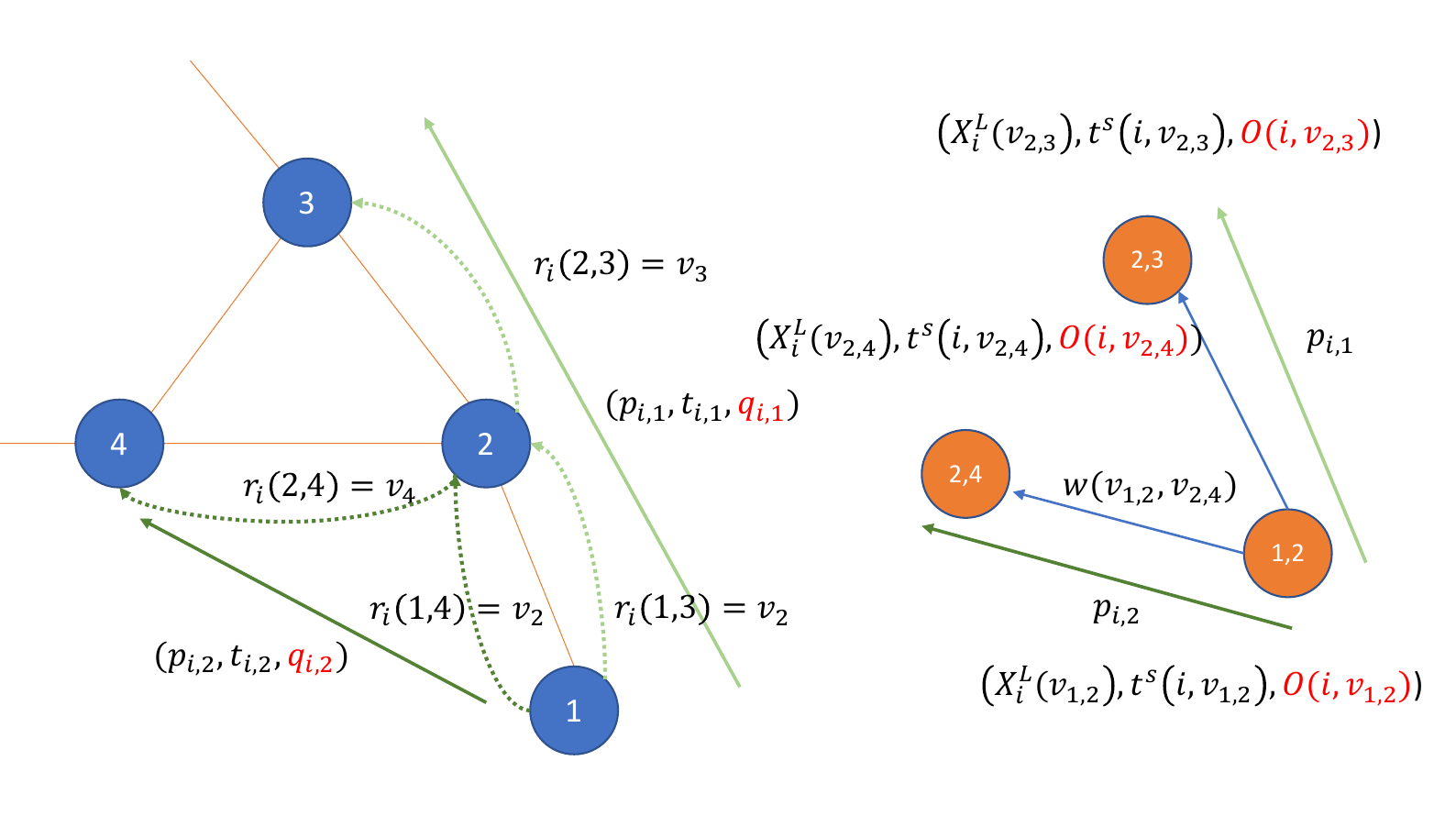}
    \caption{Comparative Example of the Problem Formulation, with Ground Truth being Red. Left: Previous works' formulation. Right: The proposed formulation}
    \label{fig:comparative}
\end{figure}
\end{inparaenum}
\begin{table}[h]
\caption{Notation Table}
\label{table:notation}
    \begin{tabular*}{0.5\textwidth}{p{0.11\textwidth}p{0.35\textwidth}}
    \toprule
    \multicolumn{2}{l}{\textbf{Network Topology Feature Notation:}}\\
    $\mathbf{G}=\{G_i\}$ & Training/testing set of Topology\\
    $G_i(V_i,E_i;X_i,Y_i)$ & Topology snapshot graph\\
    $V_i=\{v_i\}$ & Network nodes in $G_i$\\
    $n_i= |V_i|$ & Size of $G_i$\\
    $E_i=\{e_i\}$ & Communication links between network nodes\\
    $X_i =\{x(v_i)\}$ & Network node configuration of $G_i$\\
    $Y_i$ & Communication link configuration of $G_i$\\
    $c(Y(e_i))$ & Configured link capacity of $e_i \in E_i$\\
    \midrule 
    \multicolumn{2}{l}{\textbf{Network Service Feature Notation:}}\\
    $P_i=\{p_{i,j}\}$ & OD pairs in $G_i$\\
    $p_{i,j}$ & $j$-th OD-pair in $G_i$\\
    $k_i$ & Number of OD pairs in $G_i$\\
    $\mathbf{T}=\{T_i\}$ & Training/testing set of user traffic\\
    $T_i= \{t_{i,j}\}$ & User traffic matrix for all $P_i$\\
    $t_{i,j}$ & User traffic features for $p_{i,j} \in P_i$\\
    \midrule 
    \multicolumn{2}{l}{\textbf{Network Routing Feature Notation:}}\\
    $\mathbf{R}=\{R_i\}$ & Training/testing set of routing matrices\\
    $R_i$ & Routing matrices corresponds to network $G_i$\\
    $R_{i(c,d)}$ & Next-hop of current node and destination being $v_{c}, v_{d} \in V_i$\\
    \midrule 
    \multicolumn{2}{l}{\textbf{Ground Truth Notation:}}\\
    $\mathbf{Q} = \{Q_i\}$ & Training/testing set of performance matrix\\
    $Q_i$ & Performance matrix correspond to $(G_i,T_i,R_i)$\\
    $q_{i,j}$ & Performance measurement (i.e: latency) of $p_{i,j}$\\
    \midrule  
    \multicolumn{2}{l}{\textbf{Novel Feature Notation in reformulating the problem :}}\\
    $G^{\mathcal{L}}_i(V^{\mathcal{L}}_i,E^{\mathcal{L}}_i)$ & Line graph transformation of $G_i(V_i,E_i)$\\
    $\mathcal{L}^{-1}$ & Projection from representation in $G^{\mathcal{L}}_i$ to its corresponding representation in $G_i$\\
    $T^{s}_i$ & Matrix of summed traffic of $p_{i,j}$, who share the same node $v \in V^{\mathcal{L}}_i$\\
    $t^{s}(i,v)$ & Summed traffic for all $p_{i,j}$ passing through node $v \in V^{\mathcal{L}}_i$\\
    $X^{\mathcal{L}}_i(v)$ & Line graph node feature, introduced in Equ.~\ref{equ:nodeattr}\\
    $w_i$ & Line graph edge weight, introduced in Equ.~\ref{equ:edgeweight}\\
    $O(i,v)$ & node $v$'s queue occupancy state (ground truth), for $v\in V^{\mathcal{L}}_i$\\
    \bottomrule 
    \bottomrule 
    \end{tabular*}
\end{table}
\section{Methodology}
\label{sec:method}
\begin{figure}[t]
     \centering
     \begin{subfigure}[H]{0.35\textwidth}
         \centering
         \includegraphics[width=\textwidth]{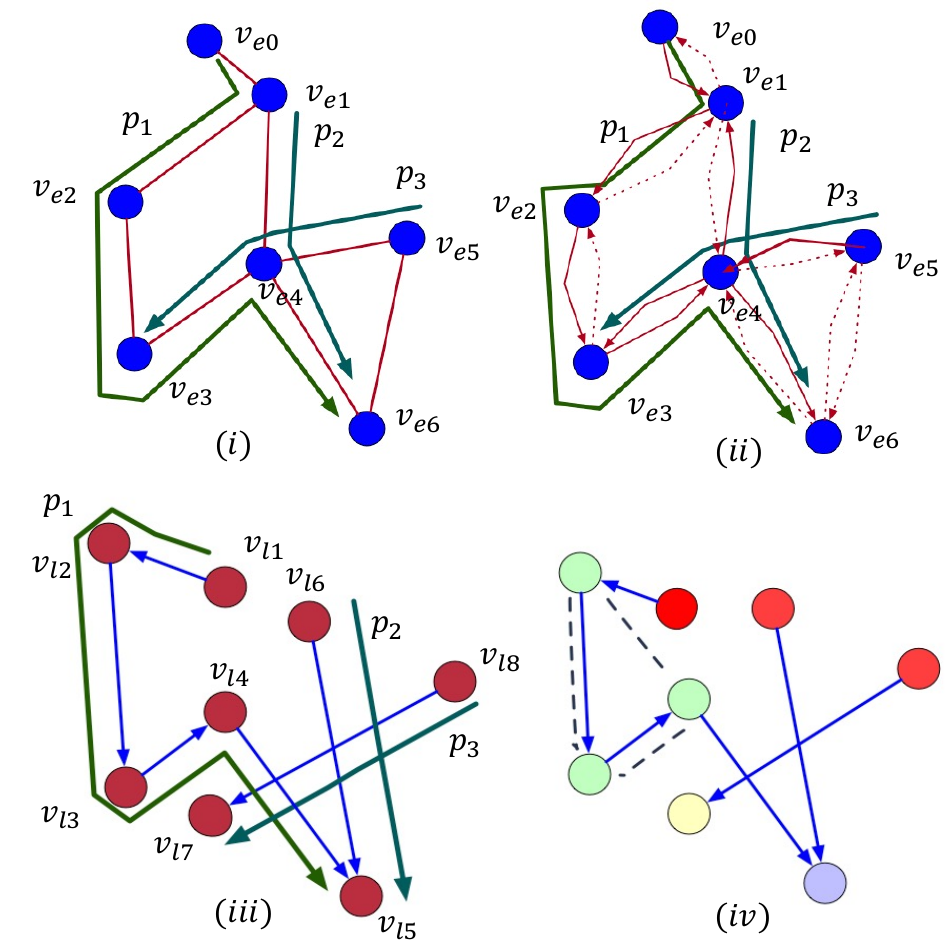}
         \caption{}
         \label{fig:Preprocessing}
     \end{subfigure}
     \hfill
     \begin{subfigure}[H]{0.5\textwidth}
         \centering
         \includegraphics[width=\textwidth]{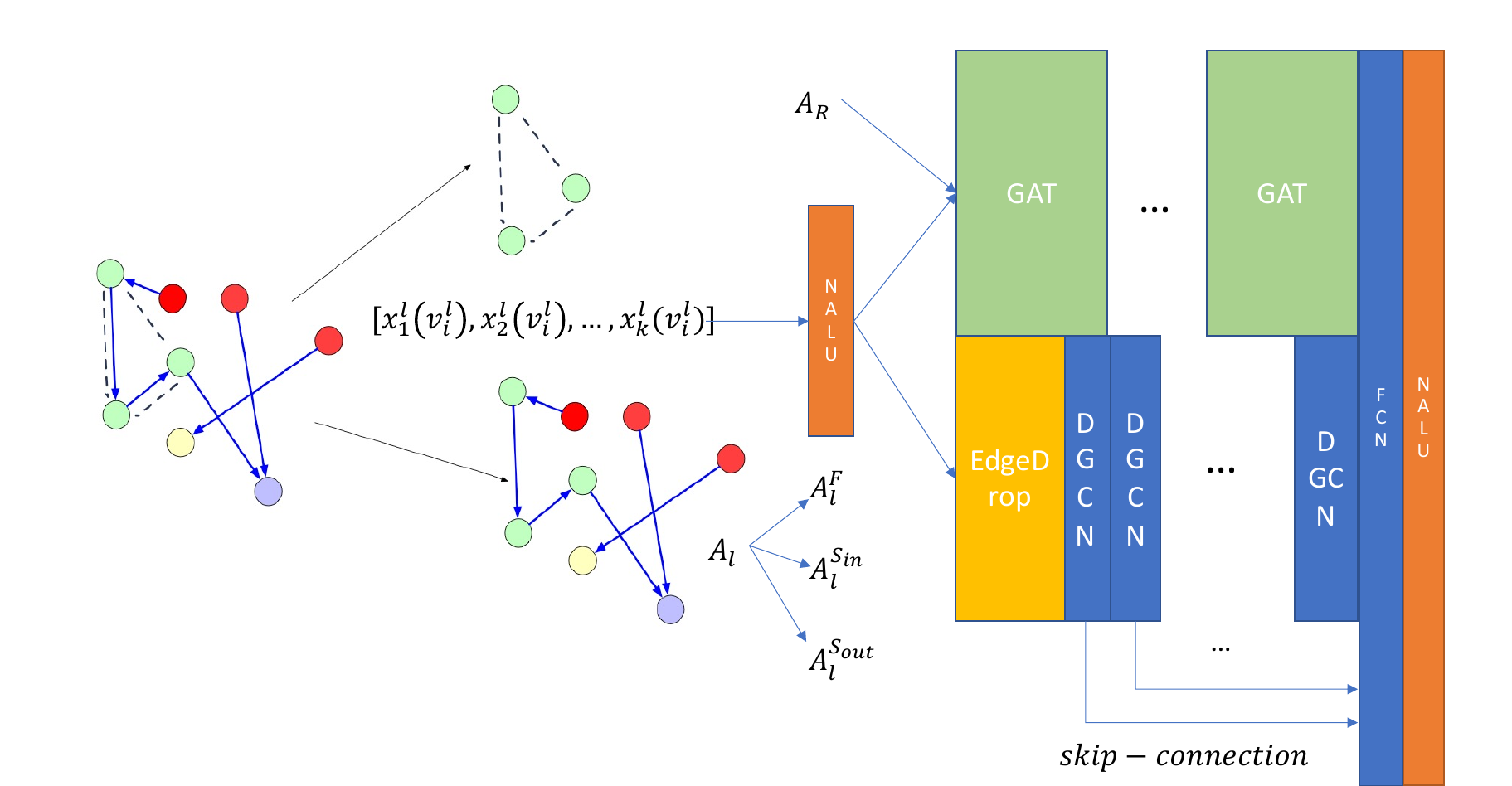}
         \caption{}
         \label{fig:model}
     \end{subfigure}
        \caption{(a) Pre-processing: Incorporate OD Information in Node's Local Structural Feature; (b) Learning and Inference: Extract Neighboring Impact and Long Range Dependencies}
        \label{fig:overview}
\end{figure}
Fig.~\ref{fig:overview} provides an overview of our proposed model for the prediction task. In the first phase (Fig.~\ref{fig:Preprocessing}), we transform the input graph $G_i$ (Fig.\ref{fig:Preprocessing} $(i),(ii)$) into the line graph $G^{\mathcal{L}}_i$ (Fig.\ref{fig:Preprocessing} $(iii),(iv)$) for the model to learn in a latent space that generalizes well on different input graph sizes. Next (Fig.~\ref{fig:model}), we separate the graph into a directed sub-graph and a un-directed sub-graph that capture the neighboring sequential relation as well as the impact of key nodes. We design the corresponding components that capture each impact and handle out-of-distribution features, so as to get consistent results across different graph sizes. Below, we discuss each component separately.

\subsection{Directed Spectral Graph Convolution} To capture the sequential relation between nodes consisting of the OD pair's trajectory, as well as the OD pair's direction, we use the architecture designed by~\cite{tong2020directed}. It is a spectral convolution block with its reception field encoded with direction information, named Directed Graph Convolution Network (DGCN). Given the processed directed $G^{\mathcal{L}}_i$, we separate the adjacency matrix into three different formations: $A^F_\ell$, $A^{S_{in}}_\ell$ and $A^{S_{out}}_\ell$, which denote first-order proximity, second-order in-proximity and second-order out-proximity, as shown in Fig.~\ref{fig:model}. Given the weighted adjacency matrix $A_\ell$ of the line graph $G^{\mathcal{L}}_i$, one can compute the aforementioned quantities:
\begin{equation}
    A^F_\ell = (A_\ell + A_\ell^T)/2,
\end{equation}
\begin{align}
\begin{split}
   A^{S_{in}}_{\ell_{(i,j)}} = \sum_k \frac{A_{\ell_{(k,i)}}A_{\ell_{(k,j)}}}{\sum_v A_{\ell_{(k,v)}}}
   \text{ and } 
    A^{S_{out}}_{\ell_{(i,j)}} =\sum_k  \frac{A_{\ell_{(i,k)}}A_{\ell_{(j,k)}}}{\sum_v A_{\ell_{(v,k)}}},&\\
   \text{ for }
   (i,j) \in V^{\mathcal{L}}_i &
\end{split}
\end{align}
We apply graph convolution to the aforementioned adjacency matrix, with adding self-loops, denoted by  $\tilde{A^F_\ell} = A^F_\ell + I$, $\tilde{A^{S_{in}}} = A^{S_{in}} + I$,  $\tilde{A^{S_{out}}} = A^{S_{out}} + I$. We concatenate (denote as $||$) each result with a train-able weight ($\alpha$, $\beta$). 
$\Theta$ is a train-able matrix. 
We define
\begin{equation}
    f_x(X,A_x) =\tilde{D_x}^{(-\frac{1}{2})}\tilde{A_x}\tilde{D_x}^{(-\frac{1}{2})}X\Theta,
\end{equation}
where  $\tilde{D_x}_{ii} = \sum_j(\tilde{A_x}_{ij})$ is 
a diagonal matrix. 
The result of each DGCN block can be formulated as:
\begin{align}
\label{equ:dgcn}
    H^{(l+1)} = ||(f_F(H^{(l)},\tilde{A^F_\ell}), \alpha f_{S_{in}}(H^{(l)},\tilde{A}^{S_{in}}), &\\
    \beta f_{S_{out}}(H^{(l)},\tilde{A}^{S_{out}})).&
\end{align}
We expect $f_F,f_{S_{in}}$ and $f_{S_{out}}$ to capture the global traffic dependency, in-traffic dependency and out-traffic dependency respectively. 
One could foresee that the deeper the model is build, such dependency can be more precisely captured. We used EdgeDrop~\cite{rong2019dropedge} as well as skip-connection to avoid over-smoothing, both of which have been found to help building deeper networks during the experimentation.

\subsection{Embedding and Readout Function with Extrapolation Ability} For handling the out of distribution attribute value per-node in larger graph, we have used NALU~\cite{trask2018neural} to replace MLP in common GNN setup. Another motivation is to build equivalence in the embedding space between a node's out-traffic attribute and its in-traffic attributes' from its neighbors. This is because the in-traffic and out-traffic of a node should remain equal, unless packet drop~happens. \cite{trask2018neural} have verified its superiority in numerical extrapolation in neural network readout.

\subsection{Routing Role Recognition based Attention}
Inspired by the work of~\cite{sun2019sinet}, identification of key nodes in the SDN is essential to estimate the whole networks performance. For this purpose, we involve role recognition~\cite{henderson2012rolx} during the pre-processing phase (denote as different colors in Fig.~\ref{fig:Preprocessing} $(iv)$). By defining $v_{s},v_{d} \in V^{\mathcal{L}}_i$ as nodes in $G^{\mathcal{L}}_i$, we consider their role to be $\mathcal{R}(v_s),\mathcal{R}(v_d)$. We define the role adjacency matrix $A_\mathcal{R}$ 
of $G^{\mathcal{L}}_i$:
\[
    A_{\mathcal{R}_{(s,d)}}= 
\begin{dcases}
    1,              & \text{if } \mathcal{R}(v_s) = \mathcal{R}(v_d), \text{ and exists } p \in P_i, \\&\text{ such that } v_s,v_d \in p  \\
    0,              & \text{otherwise}
\end{dcases}
\]
We consider the pair-wise similarity between the role of a node and its neighbors to be crucial, since an OD pair's source or destination can bring its impact towards close neighbours of nodes of interest and formulate their embedding as an ``augmented source.'' GAT is reported to be a well-suited solution to capture such pair-wise similarity impact. Thus, we apply GAT layers on node embedding between each role adjacency pair. The motivation is to capture the long-range dependency of in-trajectory of OD pair, without collecting all hops' feature in the trajectory. For the aforementioned $A_\ell$, one can consider that denser connected nodes in $G^{\mathcal{L}}_i$ are usually congested. Thus, its state (e.g., occupancy) is dependant on its neighbors' states. For less connected nodes, one can infer that they have few OD's pairs passing through. Thus, their states are more dependent on the few passing through OD pair's demanding traffic. This information might be submerged for those well-connected nodes in the OD pair trajectory, but more obvious on the weakly-connected nodes sharing the same trajectory. Such pair-wise state similarity forms $A_\mathcal{R}$,  which enlarges the receptive field of the model beyond topological neighbors. Our results show that this approach improves the model performance as well as it introduces generalization ability when training and inference takes place using samples with longer chains of trajectories. 
\section{Experiments}
\label{sec:exp}

\subsection{Dataset and Baselines}
\noindent
\textbf{Dataset.} We verify the performance of the model on a simulated dataset provided by~\cite{suarez2021graph}. The training, validation and test sets are produced by a packet level simulator (OMNeT++ v5.5.1~\cite{varga2001discrete}). Though lacking of other benchmark dataset, this dataset includes a summary of patterns of real-world network topologies from The Internet Topology Zoo~\cite{6027859}. The training set includes a variety of networks sized from 25 to 50 network nodes, while the test and validation set are sized from 50 to 300 network nodes. The larger network brings the following attribute changes: \begin{inparaenum}[\itshape (1)\upshape] \item Network topology samples introducing a few longer OD pair's trajectories; \item Network topology samples introducing larger capacity attribute per link, as well as different ``next-hop'' choice encoded in 
$R_i$, for OD pairs to generate their trajectory in a different routing preference; \item Network topology samples introducing both sampling methods (1) and (2), as described above. A description of the used dataset is shown in Table~\ref{table:dataset}.\footnote{Detailed information on feature distribution as well as comparison of the training and testing dataset is given in Appendix.D. Link to the appendix same as above.}
\end{inparaenum}

\begin{table}[b]
\caption{Network Topology Graph Set Statistics}
\begin{minipage}{.4\linewidth}
 \centering 
 \label{table:dataset}
 \begin{threeparttable}
    \begin{tabular}{p{3.5cm}rr}
    \toprule
    Dataset & Training Set & Validation \& Test set \\
    \midrule
    Number of Samples & $120000$ & $3120$ \\
    Graph size  & $[25,50]$&  $[50,300]$ \\
    Average Degree (mean,\,std) & $(9.778,0.9491)$& $(9.523,1.268)$  \\
    Average Shortest Path (mean,\,std) & $(2.379,0.1372)$& $(3.384,0.456)$  \\
    Diameter (mean,\,std) & $(4.458,0.6441)$& $(6.667,1.280)$ \\
    Cluster Coefficient (mean,\,std) & $(0.131,0.0243)$& $(0.0427,0.0342)$  \\
    \bottomrule
    \end{tabular}
\end{threeparttable}
\end{minipage}
\end{table}
\noindent
\textbf{Baselines.}
We have considered several state-of-the-art models to compare our proposed model.
\begin{enumerate}
    \item \textbf{RouteNet}~\cite{rusek2020routenet} is a message-passing-based GNN solution to estimate per OD pair's latency in SDN on bipartite graph problem formulation.
    \item \textbf{RouteNet-Salzburg}~\cite{happ2021graph} is an optimization model based on RouteNet that scales to different routing policies on same sized network topologies on estimating per OD pair's latency. This is a winning solution of \cite{suarez2021graph}.
    \item \textbf{RouteNet-GA}\footnote{Code available at \url{https://github.com/ITU-AI-ML-in-5G-Challenge/PS-014.2-GNN-Challenge-Gradient-Ascent}} is a fine-tuned solution based on RouteNet with better generalised performance on various OD pair trajectory length, this is a winning solution of \cite{suarez2021graph}.
    \item \textbf{DGCN}~\cite{tong2020directed} is a graph spectral convolution solution generalized for directed graph scenario. In our experiment, we apply NALU as its embedding and readout function, as well as same hyper-parameters and pre-processing scheme as the proposed model for a fair comparison.
    \item \textbf{Scalable-RouteNet}~\cite{ferriol2021scaling,RoutenetErlang}\footnote{We report their published results, however, we have not been given access to their implementation to reproduce their results.} is a RouteNet based improved solution to estimate per OD pair's latency in size-variant SDN on tripartite graph problem formulation.
\end{enumerate}
\subsection{Experiment Settings}
We have implemented our proposing solution with~\texttt{Keras} as well as \texttt{tf-geometric}~\cite{hu2021efficient}. Furthermore, our solution requires no-GPU, whereas for the other benchmark solutions that do require GPU, we have been training using one~\texttt{Nvidia Geforce 1080 Ti} and  \texttt{GNU-Parallel} for resource management. For training the comparative benchmark models, we used the recommended parameter sets reported by the authors, but with the unified learning rate $r = 10^{-3}$, epoch number $= 250$, epoch size $= 4000$, loss function as \textit{MAPE} as well as input feature space for a fair comparison. We train five independent random seeds for each model and report the averaged solution. The proposed method, based on aforementioned preprocessing and formulation, are taking the $G^{\mathcal{L}}_i(V^{\mathcal{L}}_i,E^{\mathcal{L}}_i)$ as well as node feature $X^{\mathcal{L}}_i(v)$ and edge weight $w_i$ in its input layer, with a $32$-dim output layer for each node embedding in $V^{\mathcal{L}}_i$. A detailed list of training parameters is published on \url{https://github.com/bluelancer/GNNET.git}.

\subsection{Experiment Result}
\begin{figure*}[t]
     \centering
     \begin{subfigure}[t]{0.32\textwidth}
         \centering
         \includegraphics[width=\textwidth]{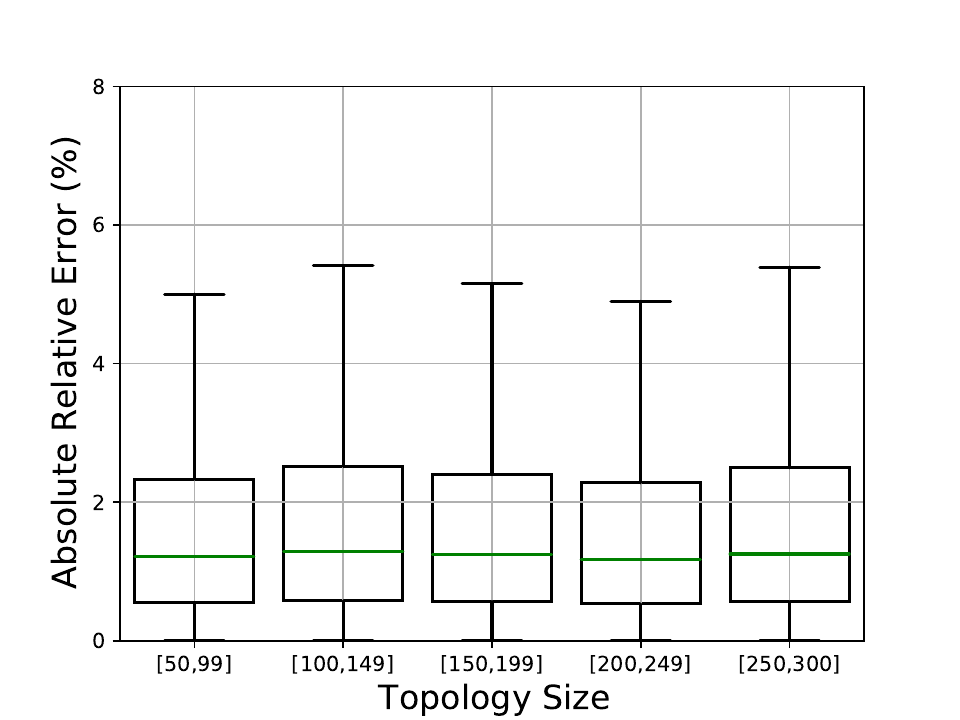}
         \caption{Proposed Model}
         \label{fig:size_result_proposed}
     \end{subfigure}
     \hfill
     \begin{subfigure}[t]{0.32\textwidth}
         \centering
         \includegraphics[width=\textwidth]{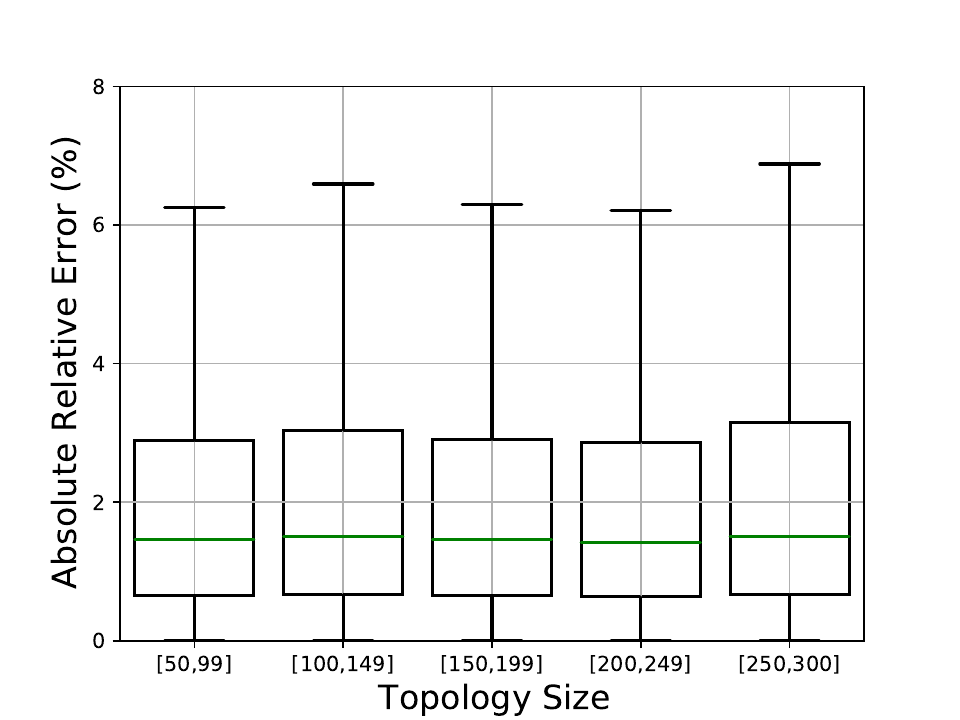}
         \caption{DGCN}
         \label{fig:size_result_dgcn}
     \end{subfigure}
     \hfill
     \begin{subfigure}[t]{0.32\textwidth}
         \centering
         \includegraphics[width=\textwidth]{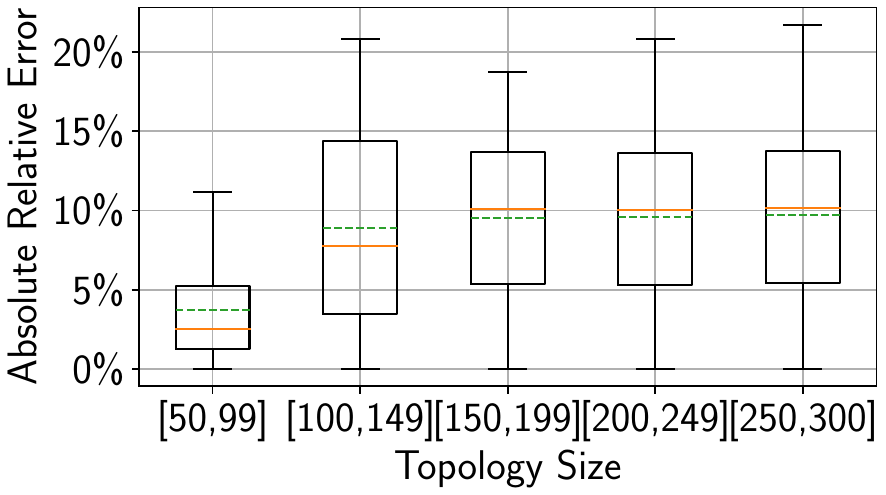}
         \caption{Scalable-RouteNet\footnotemark[\value{footnote}]}
         \label{fig:size_result_scalable_routenet}
     \end{subfigure}
     \hfill
     \begin{subfigure}[t]{0.32\textwidth}
         \centering
         \includegraphics[width=\textwidth]{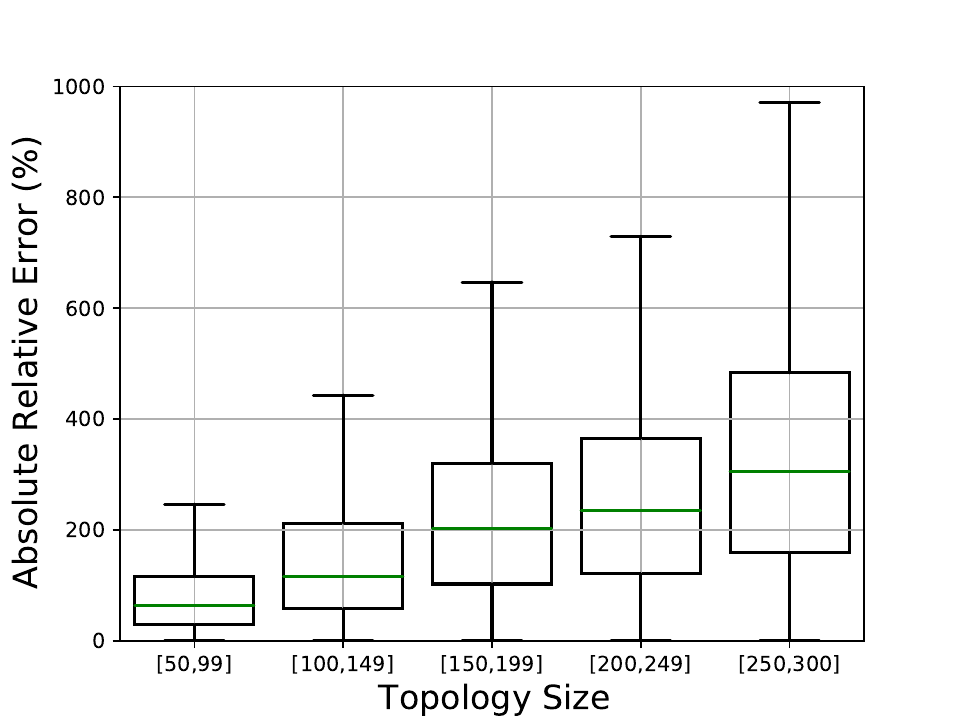}
         \caption{RouteNet}
         \label{fig:size_result_routenet}
     \end{subfigure}
     \hfill
     \begin{subfigure}[t]{0.32\textwidth}
         \centering
         \includegraphics[width=\textwidth]{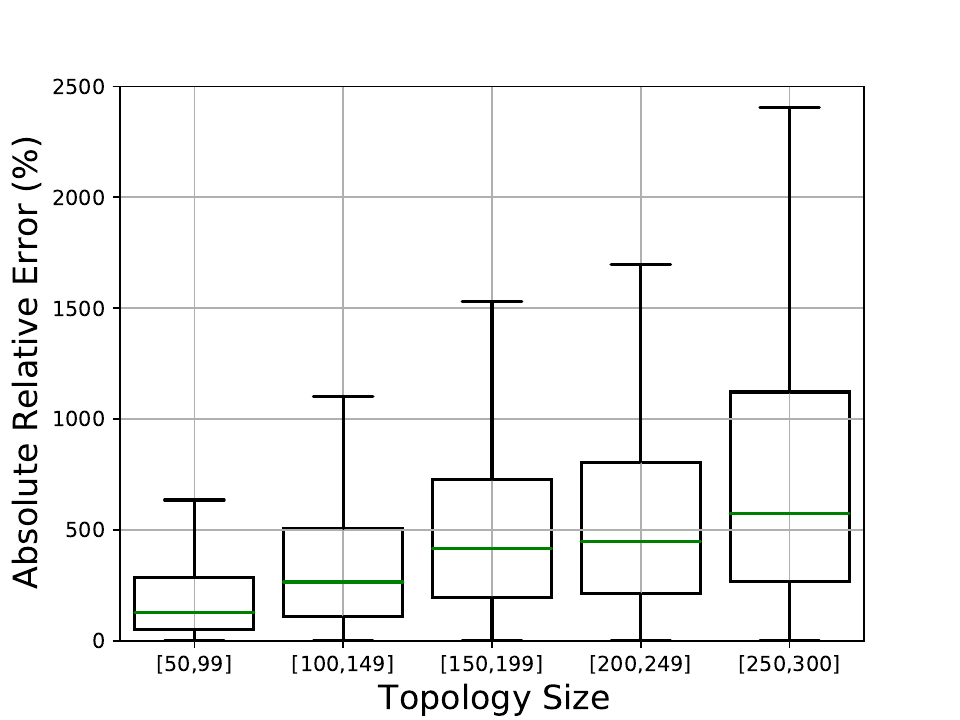}
         \caption{RouteNet-GA}
         \label{fig:size_result_routenet_ga}
     \end{subfigure}
     \hfill
     \begin{subfigure}[t]{0.32\textwidth}
         \centering
         \includegraphics[width=\textwidth]{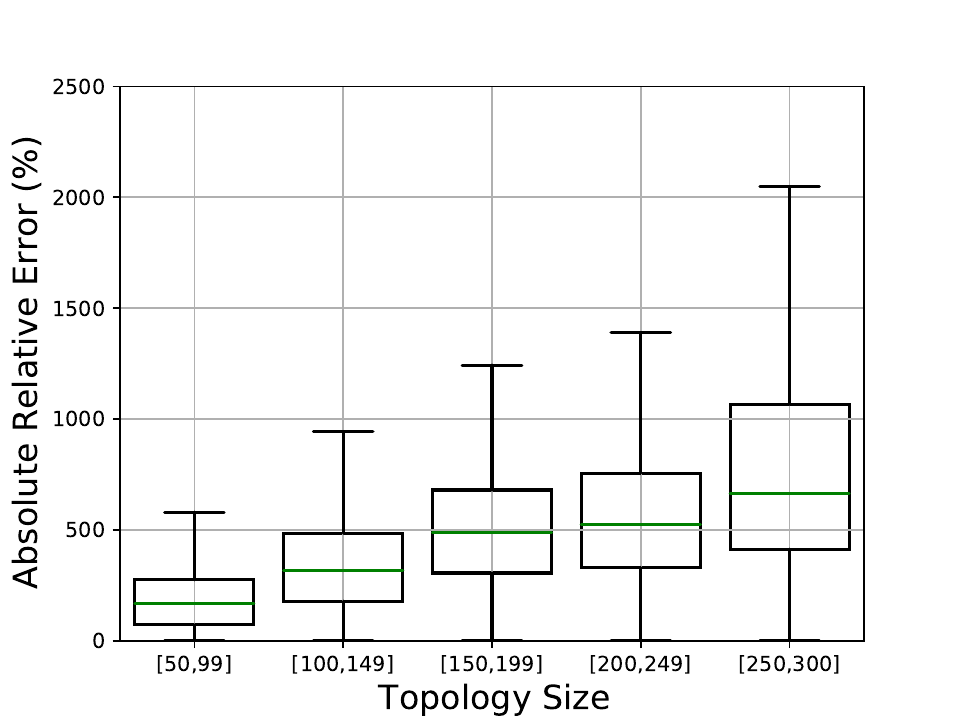}
         \caption{RouteNet-Salzburg}
         \label{fig:size_result_routenet_sal}
     \end{subfigure}
      \caption{Model Generalization Ability Towards Open-World Input}
      \label{fig:Generalization}
\end{figure*}

\begin{figure}[t!]
    \centering
    \includegraphics[width=0.4\textwidth]{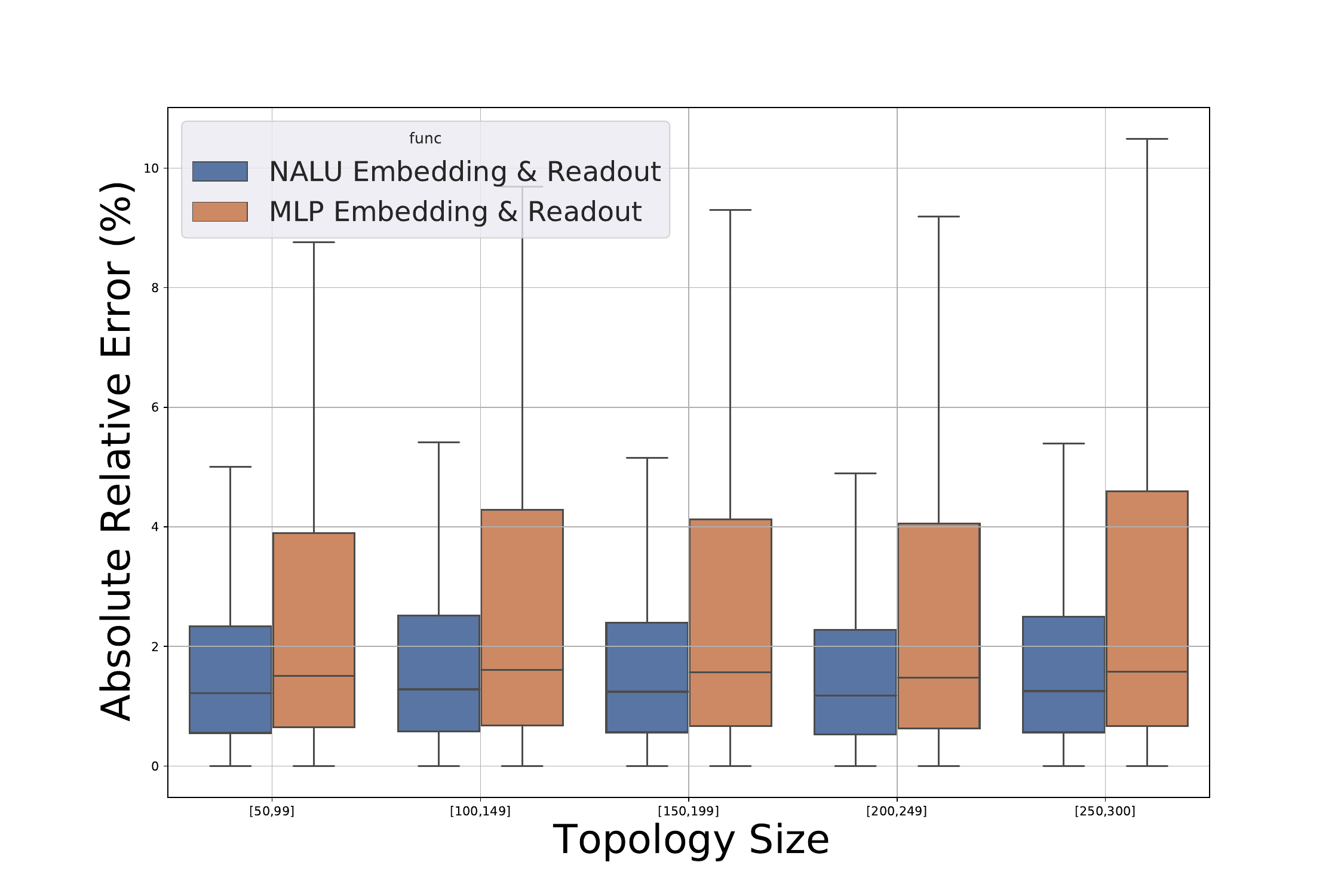}
    \caption{Ablation Study of the Proposed Model w.r.t NALU and MLP Embedding \& Readout Function}
    \label{fig:ablation}
\end{figure}

\begin{figure}[t!]
    \centering
    \includegraphics[width=0.4\textwidth]{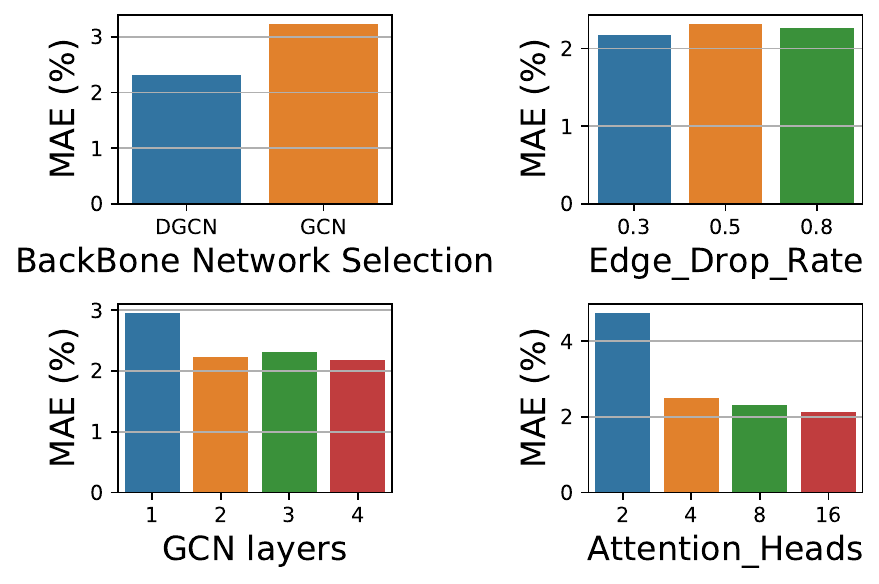}
    \caption{Sensitivity Study on the criteria of Mean Absolute Relative Error (MAE) w.r.t a Selection of Most Performance-affecting Hyper-parameter}
    \label{fig:sensitive}
\end{figure}

We evaluate the performance of our model in the following experiments: 
\begin{inparaenum}[\itshape (1)\upshape] 
\item Table~\ref{table:all_result} reports the overall performance of all the test set and we note their difference in MAPE through $Delta$ ; \item Fig.~\ref{fig:Generalization} reports model performance on each interval of size samples from the test set; \item Table~\ref{table:speed} reports mean inference speed on a selection of certain sized samples from test set. Here we are only concerned  with the time interval of inference, without any pre/post-processing time, as they can be parallelized in an actual application. We do not include results from \textbf{Scalable-RouteNet} due to the use of different hardware. \item Fig.~\ref{fig:ablation} reports an ablation study towards the decision of adopting NALU instead of MLP in proposed model when extrapolating towards different size of graph.
Fig.~\ref{fig:sensitive} reports a sensitivity study result on most affecting parameter and configuration in training the proposed model.
\end{inparaenum}
\begin{table}[H]
\caption{Average Performance\,($\%$)}
 \centering 
 \label{table:all_result}
 \begin{threeparttable}
    \begin{tabular}{lrr}
    \toprule
    Model & MAPE & Delta \\
    \midrule
    Our model & $\textbf{2.317}$   &  $-$ \\
    RouteNet & $274.5$& $272.2$  \\
    RouteNet-Salzburg & $596.8$& $594.5$  \\
    RouteNet-GA & $557.8$& $555.5$  \\
    DGCN & $ 2.935$& $0.618$  \\
    Scalable-RouteNet\footnotemark & $10$&~$7.683$  \\
    \bottomrule
    \end{tabular}
\end{threeparttable}
 \caption{Inference Speed\,($ms$) per Topology vs.\ Topology Size}
 \centering 
 \label{table:speed}
 \begin{threeparttable}
    \begin{tabular}{lrrrr}
    \toprule
    Topology Size & $50$ & $100$ & $200$ & $300$\\
    \midrule
    Our model & $161.5$&  $373.3$ &  $1115$&  $2718$ \\
    RouteNet & $485.3$& $1972$&  $6084$&  $21264$\\
    RouteNet-Salzburg &~$0.41$\,hr &~$0.4$\,hr &~$0.39$\,hr & ~$0.44$\,hr \\
    RouteNet-GA & $1363$& $4786$&  $19913$&  $71767$ \\
    DGCN & $\textbf{105.5}$& $\textbf{220.3}$&  $\textbf{861.9}$&  $\textbf{1770}$  \\
    \bottomrule
    \end{tabular}
\end{threeparttable}
\end{table}
\footnotetext{The reported result is referenced from \cite{ferriol2021scaling}}

Based on the above results, one can see that our model outperforms all baselines in terms of accuracy, inference speed and generalization ability to open-world input. 
For Table~\ref{table:all_result}, we observed that our model outperforms in terms of accuracy. We consider this verifies the validity of both our modified problem formulation, and the superiority of the proposed model. One can observe that other well-studied GNN benchmark (i.e:DGCN) can also outperform greatly against the carefully designed task-specific model, which can be concluded into the effect of problem formulation.\\
For Table~\ref{table:speed}, one can conclude that our model is faster than most task-specific benchmark models (i.e: Routenet, RouteNet-Salzburg, and RouteNet-GA). The proposed model's inference time remains tolerable for SDN routing estimation component, even for the largest network we have (300 nodes). DGCN performs marginally faster as it is one of the building blocks of the proposed model, while the proposed model is more accurate and has a more stable generalization performance against open-world input.\\
For Figure~\ref{fig:Generalization}, we conclude our proposed solution performs best in generalization ability to open-world input. Note that comparing solely DGCN and our model, we could see $A_\mathcal{R}$ not only introduces more accurate estimation, but also increases the performance stability. 
We believe that this improvement is due to introducing role inference in our model, 
which provides an easy way to diffuse long-range dependencies.\\
For Figure~\ref{fig:ablation}, we verified the benefit that we bringing in NALU instead of MLP in common GNN architecture, we observed using NALU obtains marginal gain on median accuracy but more on variance of the accuracy. This verified the conclusion from \cite{trask2018neural} that NALU assists in modeling numerical extrapolation in larger routing network samples.  

\section{Conclusion}
\label{sec:conclusion}
The paper offers the following contributions from the state of art: \begin{inparaenum}[\itshape (i)\upshape] \item We proposed a novel formulation of the latency prediction task in SDN. \item We proposed a solution that incorporates domain knowledge in SDN optimization aimed at an open-world learning setup. \item Our proposed model achieves better accuracy, inference speed and generalization ability beyond state of the art, all while using less computational resources.\end{inparaenum}
A limitation in our approach is that we do not incorporate the scenarios of peak hours, where unseen amount and duration of burst traffic triggers the network orchestration scheme, which eventually results in a dynamic-routing topology. Generalization to dynamic and size-variant graphs with GNN models remain to be studied in future work.

\section*{Acknowledgement}
This work was partially supported by the Wallenberg AI, Autonomous Systems and Software Program (WASP) funded by the Knut and Alice Wallenberg Foundation. We thank \Dr Pedro Batista and \Dr Alessandro Previti from Ericsson Research for the insightful discussions.
\medskip
\vspace*{-0.5cm}
\bibliographystyle{IEEEtran}
\bibliography{reference}
\end{document}